\documentclass[usenatbib]{mn2e}
\usepackage{graphicx}
\usepackage{mathptmx}
\def\HI{{\sc Hi}}
\def\HII{{\sc Hii}}
\def\SNR(#1.#2)#3(#4.#5){{G#1${\cdot}$#2$#3$#4${\cdot}$#5}}
\title[Re-identification of \SNR(35.6)-(0.4) as a SNR]{Re-identification of
\SNR(35.6)-(0.4) as a supernova remnant}

\label{firstpage}

\author[Green]{D.~A.~Green,\thanks{email: {\tt dag@mrao.cam.ac.uk}}\\
 Astrophysics Group, Cavendish Laboratory, 19 J.~J.~Thomson Avenue,
 Cambridge CB3 0HE}

\date{Accepted ---; received ---; in original form ---}

\pagerange{\pageref{firstpage}--\pageref{lastpage}}

\pubyear{2009}
\begin{document}

\maketitle

\begin{abstract}
\SNR(35.6)-(0.4) is an extended radio source in the Galactic plane which has
previously been identified as either a supernova remnant or an {\HII} region.
Observations from the VLA Galactic Plane Survey at 1.4~GHz with a resolution of
1~arcmin allow the extent of \SNR(35.6)-(0.4) to be defined for the first time.
Comparison with other radio survey observations show that this source has a
non-thermal spectral index, with $S \propto \nu^{-0.47 \pm 0.07}$.
\SNR(35.6)-(0.4) does not have obvious associated infra-red emission, so it is
identified as a Galactic supernova remnant, not an {\HII} region. It is
$\approx 15 \times 11$~arcmin$^{2}$ in extent, showing partial limb
brightening.
\end{abstract}


\section{Introduction}

There are over two hundred supernova remnants (SNRs) identified in the Galaxy
\citep{2004BASI...32..335G}. But current catalogues of SNRs are incomplete, not
only due to selection effects, but also because some SNRs have been
mis-identified in the past, when only limited quality observations were
available. As an example, \cite{2008ApJ...680L..37G} have recently discussed
the nature of \SNR(350.1)-(0.3). This radio source in the Galactic plane had
been included in catalogues of Galactic supernova remnants (SNRs) -- e.g.\
\cite{1976MNRAS.174..267C} and \cite{1984MNRAS.209..449G} -- following its
identification as a SNR by \cite{1975AuJPA..37....1C}. Subsequently it was
removed from later catalogues of SNRs \citep{1991PASP..103..209G}, following
the discussions of \cite{1986A&A...162..217S}, who concluded that it was not
possible to define the nature of this source from the then available
observations. \cite{2008ApJ...680L..37G} have presented new observations of
\SNR(350.1)-(0.3), including {\HI} absorption observations which show it is
Galactic, and conclude that it is after all a young and luminous SNR (although
its structure is unusual compared with other Galactic SNRs).

Prompted by this re-assessment, I present here a discussion of another extended
Galactic radio source, \SNR(35.6)-(0.4), which was included in early SNR
catalogues, but was then removed following its identification as a thermal
source. Since \SNR(35.6)-(0.4) is included in the VLA Galactic Plane Survey
(VGPS, \citealt{ 2006AJ....132.1158S}), which has a resolution of 1~arcmin, it
is now possible to better define its extent and morphology. Knowing
\SNR(35.6)-(0.4)'s extent, it is also possible to determine its radio spectrum
from available radio surveys that have lower resolution than the VGPS. Moreover
IRAS survey data can be used to distinguish between thermal or non-thermal
Galactic sources. Here I present and discuss the available radio and infrared
survey observations of \SNR(35.6)-(0.4), and conclude that it is indeed a SNR,
not an {\HII} region.

\section{Background}

\SNR(35.6)-(0.4) is Galactic radio source detected in several early radio
surveys \citep{1969AuJPh..22..121B, 1970A&AS....1..319A}, and listed as a SNR
in several early catalogues \citep{1970AuJPh..23..425M, 1971AJ.....76..305D,
1972A&A....18..169I, 1979AuJPh..32...83M}\footnote{Note that this source is
mis-labelled as \SNR(35.5)-(0.0) in Table~1 of \cite{1970AuJPh..23..425M}.}.
The extent of the source was given as $\sim 10$~arcmin, although this was not
well defined by the available observations, as they had resolutions of only
about 7~arcmin at best.

Subsequent radio observations, with somewhat better resolution, provide
contradictory information about this source. Both \cite{1974A&A....32..375V}
and \cite{1975AJ.....80..437D} derive non-thermal spectral indices for this
source, and consequently identify it as a SNR. However, it is difficult to
determine accurate flux densities for this source, since it lies on a ridge of
emission linking it to a brighter region of thermal radio emission
\SNR(35.5)-(0.0) (see below). It is also difficult to compare flux densities
derived from surveys with different resolutions when the extent of the source
is not well known. On the other hand \cite{1975AuJPA..37...57C} -- in a study
of `Observations of radio sources formerly considered as possible supernova
remnants' -- derive a flat radio spectral index of $\alpha \approx 0.04$ (using
the convention that flux density scale with frequency as $S \propto
\nu^{-\alpha}$) from 408-MHz and 5-GHz observations with similar resolution
($\approx 3$ and $4$~arcmin respectively). They regarded \SNR(35.6)-(0.4) as an
{\HII} region, not a SNR, and argue that the available upper limit in radio
recombination lines (RRLs) did not preclude an {\HII} region identification.
(Note that although \cite{1972AuJPh..25..539D} appear to report a RRL detection
from \SNR(35.6)-(0.4) and an upper limit from \SNR(35.5)-(0.0), the labels of
these objects had erroneously been swapped, see \citealt{1973AuJPh..26..267D}).
Hence, \SNR(35.6)-(0.4) was not included in the SNR catalogue of
\cite{1976MNRAS.174..267C}, nor in the first version of my catalogue
\citep{1984MNRAS.209..449G}.

\begin{figure}
\centerline{\includegraphics[width=6.5cm]{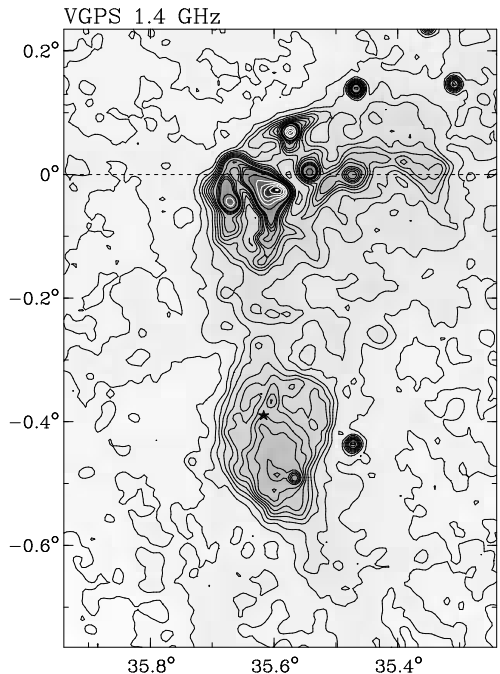}}
\caption{VGPS image of \SNR(35.6)-(0.4) and its surroundings at 1.4~GHz with a
resolution of 1~arcmin, in Galactic coordinates. The contour levels are (black)
every 2~K in brightness temperature up to 40~K, then (white) every 10~K. The
brighter region of emission towards to the top of the plot is the thermal
source \SNR(35.5)-(0.0). The star indicates the position of PSR 1855$+$02. The
dashed lines marks the Galactic plane where $b=0^\circ$.}\label{f:vgps}
\end{figure}

\begin{figure}
\centerline{\includegraphics[width=6.5cm]{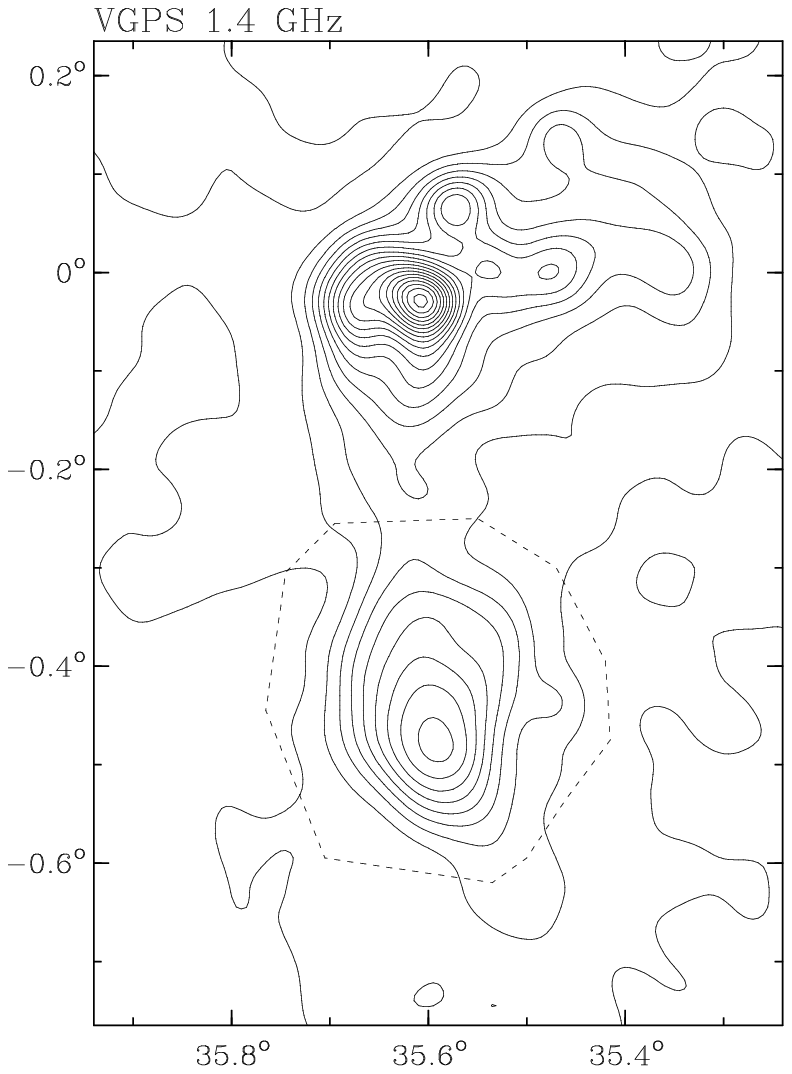}}
\medskip
\centerline{\includegraphics[width=6.5cm]{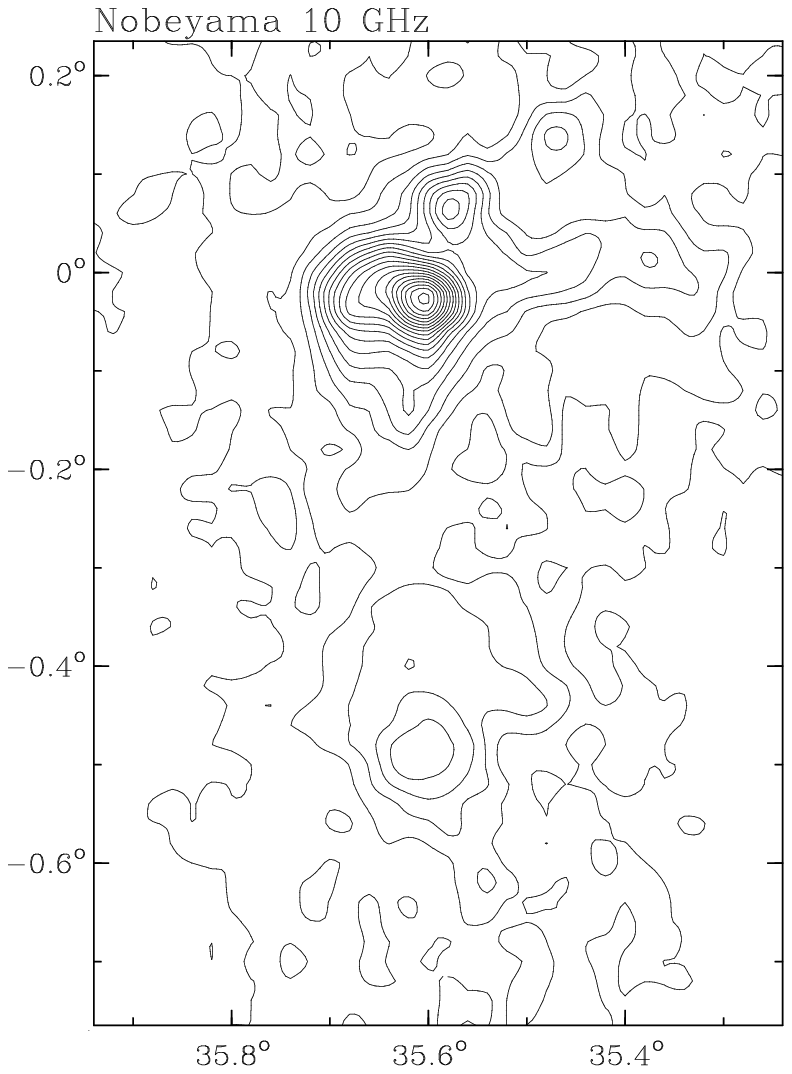}}
\caption{Images of \SNR(35.6)-(0.4) and its surroundings, in Galactic
coordinates. (Top) VGPS at 1.4~GHz smoothed to a resolution of 3~arcmin. The
contour levels are 2.2~K in brightness temperature. A polygon used to derived
an integrated from density for \SNR(35.6)-(0.4) is shown (see text). (Bottom)
Nobeyama survey at 10~GHz from Handa et al.\ (1987) with a resolution of
3~arcmin. The contour levels are every 0.033~K in brightness
temperature.}\label{f:alpha}
\end{figure}

\begin{figure}
\centerline{\includegraphics[width=6.5cm]{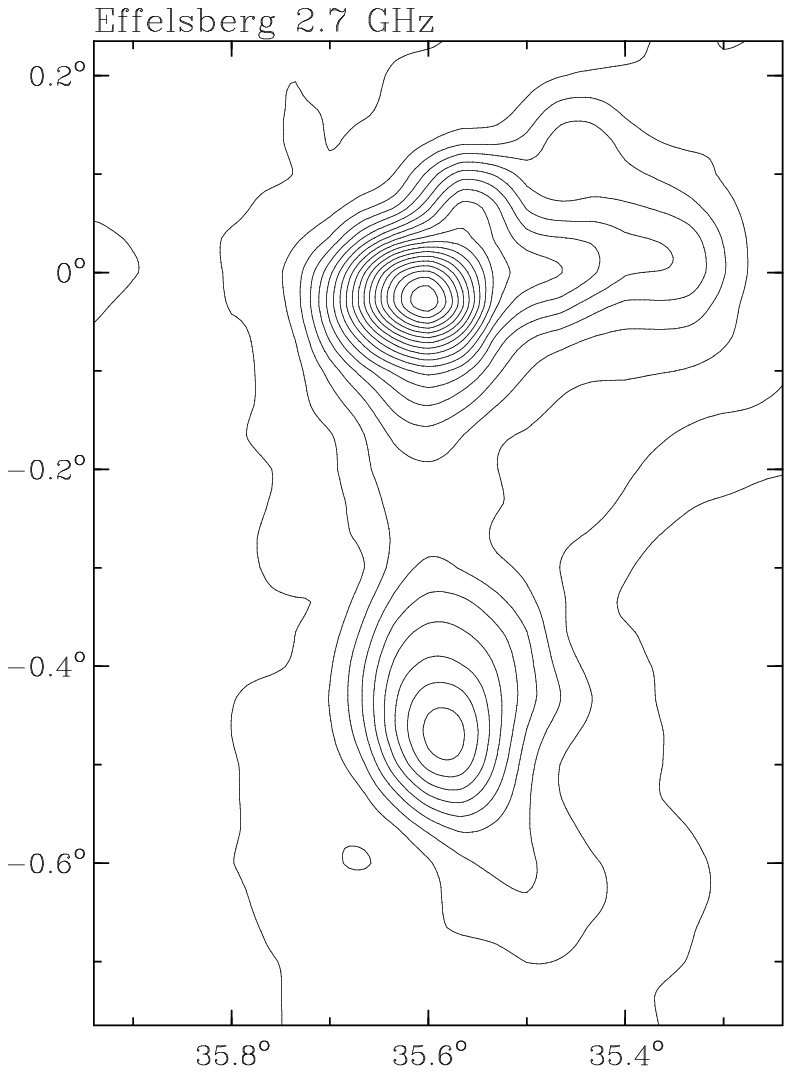}}
\medskip
\centerline{\includegraphics[width=6.5cm]{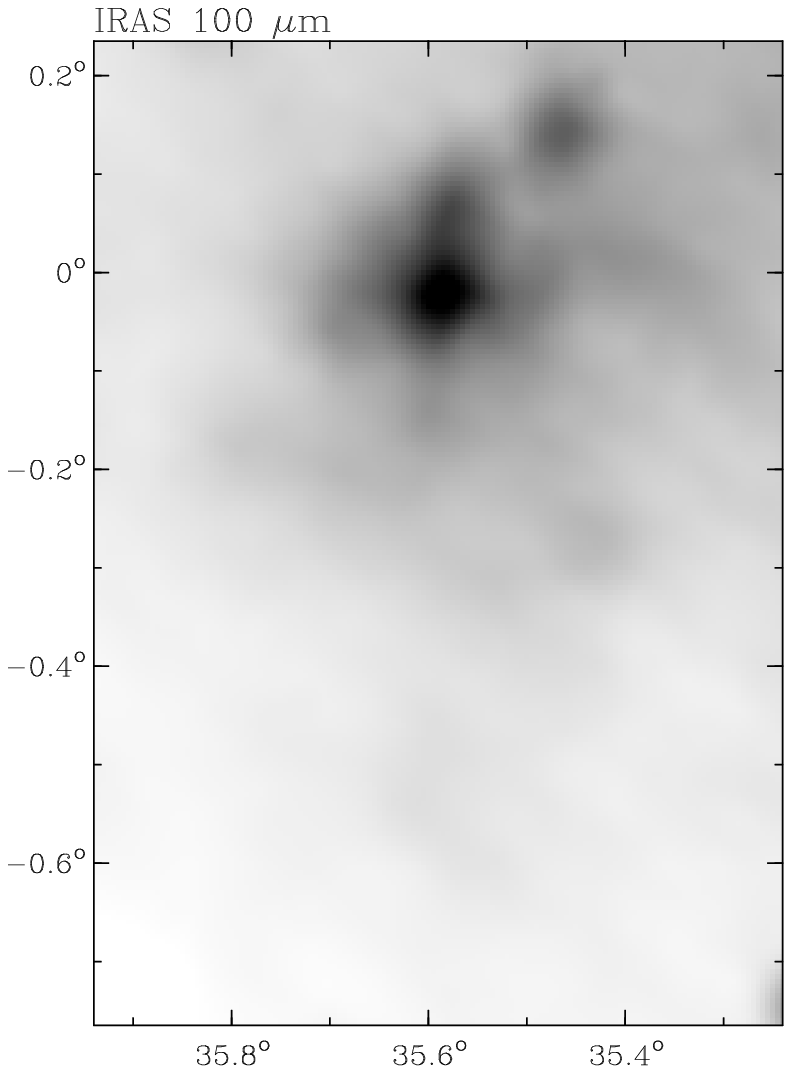}}
\caption{Images of \SNR(35.6)-(0.4) and its surroundings, in Galactic
coordinates. (Top) Effelsberg 2.7-GHz image from Reich et al.\ (1984) with a
resolution of 4.3~arcmin. The contour levels are every 0.45~K in brightness
temperature. (Bottom) IRIS 100-$\mu$m image with a resolution of 4.3~arcmin.
The greyscale is 400 to 3000~MJy~sr$^{-1}$.}\label{f:iris}
\end{figure}

\cite{1977A&A....55...11A} also derived a steep non-thermal spectral index of
\SNR(35.6)-(0.4), and concluded that it is a SNR. They also found faint
polarisation towards it at 5~GHz, but noted that it was not clear if this was
associated with the source, or from the Galactic background. More recently
\cite{1989ApJS...71..469L} report a RRL line detection towards $l=35\fdg588$,
$b=-0\fdg489$, and consequently this source -- labelled \SNR(35.6)-(0.5) -- was
included as an {\HII} region in later studies, e.g.\
\cite{1990ApJ...352..192K}, who made a marginal detection of {\HI} absorption
towards it. \SNR(35.6)-(0.5) is also discussed by \cite{1993ASPC...35..419P},
who describe it as an SNR--{\HII} region complex, based on
\citeauthor{1972AuJPh..25..539D}'s reported RRL detection, and
\citeauthor{1977A&A....55...11A} report of non-thermal emission.
\citeauthor{1993ASPC...35..419P} present some further RRL detections towards
both \SNR(35.6)-(0.5) and \SNR(35.5)-(0.0), and propose an association between
the supposed SNR in the region and the rather old pulsar PSR 1855$+$02.

\section{Observations}

\SNR(35.6)-(0.4) is included in the region covered by the VGPS, with a
resolution of 1~arcmin at 1.4~GHz, see Fig.~\ref{f:vgps}. This provides the
highest resolution radio image of the source, which is shown as a partially
limb-brightened region of emission $\approx 15 \times 11$~arcmin$^{2}$.
Fig.~\ref{f:vgps} also shows the emission from the thermal complex
\SNR(35.5)-(0.0). The two compact sources within and near \SNR(35.6)-(0.4) are
IRAS 18551$+$0159 (near $l=35\fdg47$, $b=-0\fdg44$) and IRAS 18554$+$0203 (near
$l=35\fdg56$, $b=-0\fdg49$), which are a candidate post-AGB star and a
planetary nebula respectively (see \citealt{1999ApJS..123..219C,
2005A&A...431..779J, 2006MNRAS.373...79P, 2007ApJ...669..343C,
2008ApJS..174..426K} and references therein). IRAS 18551$+$0159 and IRAS
18554$+$0203 have $\approx 0.117$ and $0.096$~Jy at 1.4~GHz
(\citeauthor{1999ApJS..123..219C}).

The region containing \SNR(35.6)-(0.4) is also included in several single dish
radio surveys, which are available electronically. Fig.~\ref{f:alpha} shows an
image of \SNR(35.6)-(0.4) and its surroundings from the 10-GHz survey by
\cite{1987PASJ...39..709H}\footnote{see also: {\tt
http://www.ioa.s.u-tokyo.ac.jp/{\char'176}handa/}}, compared with the VGPS
image at 1.4~GHz, smoothed to matched resolution of 3~arcmin. The quality of
the 10-GHz survey varies with galactic longitude, depending on observing
conditions, but \SNR(35.6)-(0.4) is a region of relatively high quality. The
contour levels in this figure have been chosen so that optically thin thermal
emission with $\alpha = 0.1$ appears the same at each frequency (i.e.\ the
spacing between the contours in brightness temperature is scaled by
$(1.42/10.5)^{2.1} \approx 0.015$).

This clearly shows that, the emission from \SNR(35.6)-(0.4) has a steeper
spectral index than $\alpha = 0.1$, as there is relatively more emission from
it at 1.4~GHz. In contrast, the emission from \SNR(35.5)-(0.0) is consistent
with thermal emission. The non-thermal spectrum of the emission from
\SNR(35.6)-(0.4) is supported by comparison with other single-dish survey data
with lower resolution that are available electronically, namely the Effelsberg
2.7-GHz survey \citep{1984A&AS...58..197R} and the Parkes 5.0-GHz survey
\citep{1978AuJPA..45....1H}.

The flux density of \SNR(35.6)-(0.4) was estimated from the smoothed VGPS
1.4-GHz and Nobeyama 10-GHz images (i.e.\ Fig.~\ref{f:alpha}). Polygons were
drawn around \SNR(35.6)-(0.4), a tilted plane was fitted to the edges of the
polygons and then removed from the image, with the remaining emission within
the polygon integrated (see \citealt{2007BASI...35...77G}). This procedure is
subjective in that the results depend on the exact choice of polygon used. In
this case the full-resolution VGPS image was used to define the polygons,
bearing in mind the lower resolution of the 10-GHz Nobeyama. Several polygons
were used, one of which is shown in Fig.~\ref{f:alpha} (note that this includes
IRAS 18551$+$0159 which is outside of \SNR(35.6)-(0.4) in the full resolution
VGPS image, Fig.~\ref{f:vgps}, but cannot be separated from \SNR(35.6)-(0.4) in
the 3~arcmin resolution images). The same polygons were used for both for the
1.4-GHz and 10-GHz images, and it was found that the derived integrated flux
densities varied by about 5~per~cent at most. The derived flux densities for
\SNR(35.6)-(0.4) are: $7.8$ and $3.1$~Jy at 1.4 and 10~GHz respectively. Taking
a cautious error of 10~per~cent in each of the flux densities, these give a
spectral index of $\alpha = 0.46 \pm 0.07$ for the radio emission from
\SNR(35.6)-(0.4).

Given the previous identifications of \SNR(35.6)-(0.4) as either a SNR or a
thermal source, infra-red observations allow further discrimination between
these possibilities. Fig.~\ref{f:iris} shows an IRAS image of \SNR(35.6)-(0.4)
at 100~$\mu$m from Improved Reprocessing of the IRAS Survey (IRIS,
\citealt{2005ApJS..157..302M}\footnote{see also {\tt
http://irsa.ipac.caltech.edu/data/IRIS/}}), compared with a radio image at
2.7~GHz from the Effelsberg survey of \cite{1984A&AS...58..197R}\footnote{see
also {\tt http://www.mpifr-bonn.mpg.de/survey.html}}, which has comparable
resolution. Fig.~\ref{f:iris} shows infra-red emission clearly associated with
\SNR(35.5)-(0.0), which is known to be thermal, but no obvious infra-red
emission from \SNR(35.6)-(0.4).

\section{Discussion and Conclusions}

From the results presented above, \SNR(35.6)-(0.4) can be identified as a SNR,
not an {\HII} region, because of its non-thermal radio spectrum and the lack of
extended infra-red emission associated with it. However, the detection of RRL
from this source by \cite{1989ApJS...71..469L} and \cite{1993ASPC...35..419P},
implies that there is some thermal emission towards this region. But, as noted
above, there are indeed two compact IRAS sources within or near
\SNR(35.6)-(0.4), with the PN IRAS 18554$+$0203 being close to the positions
where RRLs have been detected.

The VGPS observations (Fig.~\ref{f:vgps}), show that \SNR(35.6)-(0.4) has a
partially limb-brightened structure -- i.e.\ it is a `shell' remnant -- but
with differing radii in different directions, which is reminiscent of the SNR
\SNR(166.0)+(4.3) ($=$VRO 42.05.01), see \cite{1989MNRAS.237..277L}.
\SNR(35.6)-(0.4) has a 1~GHz surface brightness of $\approx 8 \times 10^{-21}$
W~m$^{2}\,$Hz$^{-1}$\,sr$^{-1}$ (for 1~GHz flux density of 9~Jy, and angular
size of $15 \times 11$~arcmin$^2$), which is close to what is thought to be the
surface brightness completeness limit for current Galactic SNR catalogues
(e.g.\ \citealt{2004BASI...32..335G}).

Inspection of the {\HI} line observations in the VGPS, smoothed to channel
resolution of $\approx 8$~km~s$^{-1}$, shows absorption towards
\SNR(35.6)-(0.4) at positive velocities, up to $\approx +55$~km~s$^{-1}$.
However, the brightness temperature of the emission from \SNR(35.6)-(0.4) is
low, $\approx 20$~K above the local background (see Fig.~\ref{f:vgps}), so that
faint absorption will not be easily seen. This absorption provides a lower
limit to the distance to \SNR(35.6)-(0.4) of $\approx 3.7$~kpc (the near
distance corresponding to $+55$~km~s$^{-1}$, for a simple `flat' rotation curve
with a constant velocity of 220~km~s$^{-1}$ and a Galactocentric radius of
8.5~kpc). In the VGPS {\HI} line observations, absorption towards the brighter
{\HII} region \SNR(35.5)-(0.0) can be seen at larger positive velocities, to at
least $+90$~km~s$^{-1}$. The recombination line velocity for this {\HII} region
is $+51$~km~s$^{-1}$ \citep{1989ApJS...71..469L}, which implies a distance of
$\approx 10.5$~kpc (i.e.\ the far distance corresponding to $51$~km~s$^{-1}$,
since {\HI} absorption is seen to larger positive velocities). If
\SNR(35.6)-(0.4) is at the same distance as \SNR(35.5)-(0.0), then its physical
size is $\approx 46 \times 34$~pc$^2$, which is consistent with the typical
physical sizes of SNRs with known distances for a surface brightness of
$\approx 8 \times 10^{-21}$ W~m$^{2}$\,Hz$^{-1}$\,sr$^{-1}$
\citep{2004BASI...32..335G}. PSR 1855$+$02 is close to the centre of
\SNR(35.6)-(0.4), and as noted above, \citeauthor{1993ASPC...35..419P} proposed
an association of it with the remnant. \cite{2005MNRAS.360..974H} derive a
distance of $\approx 8$~kpc for PSR 1855$+$02, from its observed dispersion
measure, which is in reasonable agreement with this distance for
\SNR(35.6)-(0.4). For a simple Sedov--Taylor model, with a nominal explosion
energy of $10^{44}$~J and an ambient density of 1~H atom cm$^{-3}$, this would
imply an age of about 30 thousand years for \SNR(35.6)-(0.4). PSR 1855$+$02 has
a characteristic age of $\approx 160$ thousand years
\citep{2004MNRAS.353.1311H}. Thus, it appears difficult to associated PSR
1855$+$02 with this SNR. However, given the uncertainties in both the distance
to the remnant, and the simple Sedov--Taylor model, it is difficult to say
anything definitive on the proposed pulsar--SNR association.

\section*{Acknowledgements}

The National Radio Astronomy Observatory is a facility of the National Science
Foundation operated under cooperative agreement by Associated Universities,
Inc. This research has made use of NASA's Astrophysics Data System
Bibliographic Services.

\setlength{\labelwidth}{0pt} 

\bsp

\label{lastpage}
\end{document}